\documentstyle[a4wide,12pt,epsf]{article}
\begin{document}
\begin{flushright}
\baselineskip=12pt
{SUSX-TH-01-032}\\
{RHCPP 01-03T}\\
{hep-th/0108131}\\
{July  2001}
\end{flushright}
\def\IZ{Z\kern-.4em Z}
\begin{center}
{\LARGE \bf STANDARD-LIKE MODELS 
FROM INTERSECTING D4-BRANES \\}
\vglue 0.35cm
{D.BAILIN$^{\clubsuit}$ \footnote
{D.Bailin@sussex.ac.uk}, G. V. KRANIOTIS$^{\clubsuit}$ \footnote
 {G.Kraniotis@sussex.ac.uk} and A. LOVE$^{\spadesuit}$ \\}
	{$\clubsuit$ \it  Centre for Theoretical Physics, \\}
{\it University of Sussex,\\}
{\it Brighton BN1 9QJ, U.K. \\}
{$\spadesuit$ \it  Centre for Particle Physics , \\}
{\it Royal Holloway and Bedford New College, \\}
{\it  University of London,Egham, \\}
{\it Surrey TW20-0EX, U.K. \\}
\baselineskip=12pt

\vglue 2.5cm
ABSTRACT
\end{center}
We construct a one-parameter set of intersecting D4-brane models, with six stacks, that yield the (non-supersymmetric) 
standard model plus extra vector-like matter. Twisted tadpoles and gauge anomalies are cancelled, and 
 the model contains all of the Yukawa couplings to the tachyonic Higgs doublets that are needed to generate mass terms for the fermions. 
 A string scale in the range $1-10$ TeV  and a Higgs mass not much 
greater than the current bound
  is obtained for certain values of the parameters, consistently with the 
observed values of the gauge coupling constants.

{\rightskip=3pc
\leftskip=3pc
\noindent
\baselineskip=20pt

}

\vfill\eject
\setcounter{page}{1}
\pagestyle{plain}
\baselineskip=14pt
The use of D-branes in Type II string theories has led to exciting new possibilities for constructing models with realistic gauge groups and three 
generations of chiral matter \cite{UTCA}. Since D-branes carry gauge fields localised on their world volumes, an attractive approach
 is to start from a configuration of D-branes designed to give the standard model gauge group, possibly augmented by additional $U(1)$ 
 factors, and to introduce further D-branes to produce at least the required matter content.  The  branes are located at an 
 $R^6/\IZ_N$ orbifold fixed point, to ensure ${\mathcal N}=1$ supersymmetry, and the whole system is then embedded in a global compact
  space, an orbifold or orientifold, for example. Further D-branes and/or Wilson lines as well as anti-D-branes
   are added to ensure the cancellation   of twisted tadpoles and the overall consistency of the theory. 
   This is the ``bottom-up" approach pioneered in \cite{AIQU}, who also showed that only the $\IZ_3$ point group can yield three chiral
    generations of matter. Extra vector-like matter is generic to such models, and it was shown in \cite{BKL} that this leads
 to gauge unification at an intermediate string scale between about $10^{10}$ and $10^{12}$ GeV, consistently with the measured value of
 $\sin^2 \theta_W(m_Z)$. Despite their attractions these models are not without defects. Specifically, the unavoidable
  extra D-branes needed to 
 construct the model generate additional gauge groups under which some of the standard model matter is charged. So these gauge groups
  are not hidden from the observable sector, and the absence of data for such extra gauge groups must be 
  taken as evidence against such models.
Further, some of the Yukawa terms needed to generate masses for the quarks and leptons are forbidden by surviving
 global $U(1)$ symmetries. Also, these global
 symmetries are spontaneously broken when electroweak symmetry breaking occurs, thereby generating a keV-scale axion \cite{MBH} which is 
 unambiguously excluded by axion searches.
 
A different route to the standard model was motivated by Berkooz, Douglas and Leigh's  observation \cite{BDL} that
 intersecting D-branes can give rise to chiral fermions propagating in the 
intersection of their world volumes. 
D-branes at angles afford a T-dual interpretation as D-branes with magnetic 
flux \cite{CBACHAS}, and based on that paper the stringy methods for 
what are now called ``intersecting brane models'' were developed in 
\cite{LUST}.
This last paper led Aldazabal et al \cite{AFIRU1, AFIRU2} to develop (non-supersymmetric) four-dimensional
chiral models from intersecting D4-branes wrapped on 1-cycles of a 2-dimensional torus $T^2$ sitting at a singular point 
in the transverse 4-dimensional space $B$ \footnote{ For other 
recent work on  interesecting brane models, both 
supersymmetric and non-supersymmetric, and their phenomenological 
implications, see reference \cite{INTERSECTING}.}.  The local geometry of $B$ near the singularity is modelled as $C^2/\IZ_N$. 
This gives  an ${\mathcal N}=2$ supersymmetric gauge sector. Since 
the models are generically non-supersymmetric, by virtue of their matter content, solution of the hierarchy 
problem requires a string scale of not more than $10$ TeV 
 or so, although the usual four-dimensional Planck mass can be obtained by making the volume of $B$ large enough.
  The generic
  appearance at tree level of scalar tachyons at some intersections allows the interpretation of doublets as electroweak Higgs scalars, although 
  colour-triplet 
  and/or charged singlet tachyons are potentially lethal for theories purporting to be realistic. However, provided that 
  their masses are small compared with the string scale, radiative corrections might overcome this defect. Although these
   models are non-supersymmetric, they typically contain extra matter,
   besides that in the standard model, some of which can be light. In particular there are towers of ``gonions'', massive 
   (fermionic and bosonic), vector-like
    matter   arising from excitations associated 
  with the angle at which the D-branes intersect, and which are charged with respect to the standard model gauge group. There are 
  also Kaluza-Klein and winding-mode towers of gauge bosons which might also exist below the string scale. 
  
  All of these features make intersecting brane models susceptible to experimental investigation in the near future, 
  so it is clearly desirable to construct models that are consistent with existing data.
   A first step in this programme was made in \cite{AFIRU2} where some three-generation models were constructed using five stacks of
   D4-branes, each stack wrapping a different 1-cycle of $T^2$. 
   Note that two stacks are different if their 1-cycles intersect, or if they are parallel but not overlapping.
      The stacks sit
    at a singular point of the transverse space $B$, taken to be $(T^2)^2/\IZ_3$. 
 Our objective in this paper is to 
construct models which besides satisfying the technical constraints of Ramond-Ramond 
tadpole and anomaly cancellation also satisfy more of the phenomenological constraints. Specifically, we require that the models allow 
the (renormalizable) Yukawa couplings needed to generate mass terms for 
all standard model matter. The gauge couplings in 
  these models are {\it not} unified at the string scale; they depend upon the wrapping numbers of the particular stack of D4-branes
   with which they are associated. So we also require that these values 
are achieved at a string scale of not more than about $10$ TeV using 
  renormalization group equations starting from the measured values of the gauge coupling strengths at the weak scale.
    We find that these 
   constraints cannot be simultaneously satisfied using at most five stacks of D-branes, but that certain six-stack models can do so.

As in \cite{AFIRU2}, we take the space $B$ transverse to the D4-branes to be $C^2/\IZ_3$, where the point group generator
$\theta$ of $\IZ_3$ acts on the two complex coordinates of $B$ with twist vector $v=\frac{1}{3}(1,-1)$. 
We too choose the first stack 
to have $N_1=3$ D4-branes  and to be sitting at a singular point of $B$ with wrapping numbers
 $(n_1,m_1)=(1,0)$. In general the integers $n_a$ and $m_a$ count the number of 
 times that the 1-cycle 
 wrapped by the stack $a$ wraps the two basis 1-cycles defining the torus $T^2$; there is no loss of generality in this choice of  
wrapping numbers for the first stack.  $\theta$ 
  is embedded in the stack as $\gamma_{\theta}=\alpha ^p {\bf I}_3$ which generates a $U(3) \supset SU(3)_c$ gauge group.
Here $\alpha = e^{2\pi i/3}$ and $p=0,1 \quad {\rm or} \quad 2 $. As in \cite{AFIRU2}, 
we also choose the second stack to generate
  a $U(2) \supset SU(2)_L$ gauge group by taking $N_2=2$ D4-branes and embedding $\theta$ as $\gamma_{\theta}=\alpha ^q {\bf I}_2$, 
  where $q \neq p \bmod 3$. The wrapping numbers are $(n_2, 3)$, so that the $(12)$ intersection produces 
  $I_{12} \equiv n_1m_2-n_2m_1=3$ quark doublets $Q_L$. Note that, unlike in the bottom-up approach \cite{AIQU}, 
the number of generations is not determined by the choice of the point group $\IZ_3$.  Since further non-abelian gauge groups are not required, all remaining stacks have just one D4-brane. 
  Without loss of generality these remaining stacks are split into three sets $I,J$ and $K$ having  $\gamma_{\theta}=\alpha ^q, 
  \alpha ^r$ and $\alpha ^p$ respectively, where $p,q,r$ are a permutation of $0,1,2$. The wrapping numbers
   for the stack $i \in I$ are $(n_i,m_i)$,
  and similarly for $j \in J$ and $k \in K $. Table ~1 summarizes the assignments.
\begin{table}
\begin{center}
\begin{tabular}{|c|c|c|c|} \hline \hline
Stack $a$ & $N_a$ & $(n_a,m_a)$ & $\gamma_{\theta}$ \\
\hline \hline
1 & 3 & $(1,0)$ & $\alpha ^p {\bf I}_3$ \\
\hline
2 & 2 & $(n_2,3)$ & $\alpha ^q {\bf I}_2 $\\
\hline
$i \in I$ & 1 & $(n_{i},m_{i})$ & $\alpha ^q $\\
\hline
$j \in J $& 1 & $(n_{j},m_{j})$ & $\alpha ^r$ \\
\hline
$k \in K$ & 1 & $(n_{k},m_{k})$ & $\alpha ^p$ \\
\hline \hline
\end{tabular}
\end{center}
\caption{Multiplicities, wrapping numbers and Chan-Paton phases for the D4-brane stacks, 
($p\neq q \neq r \neq p$).}
\end{table}
The gauge group resulting from these stacks of branes is at least 
  \begin{equation}
  \prod_a U(N_a)=U(3) \times U(2) \times \prod_i U(1)_i \times \prod_j U(1)_j \times \prod_k U(1)_k
  \label{321p}
  \end{equation}
When $n_a$ and $m_a$ have a greatest common divisor greater than 1 or one 
of the $n_a$ or $m_a$ is zero, there may be additional gauge group  factors.
The wrapping numbers are constrained by the requirements of twisted tadpole cancellation, which for the system described 
above gives
\begin{eqnarray}
3+\sum_{k}n_{k}=2n_2+\sum_{i}n_{i}=\sum_{j}n_{j} \label{n}\\
\sum_{k}m_{k}=6+\sum_{i}m_{i}= \sum_{j}m_{j} \label{m}
\end{eqnarray}

The spectrum of these models is generically chiral, so potentially there are gauge anomalies.
 In general, tadpole cancellation is sufficient to ensure cancellation of the cubic non-abelian anomalies, which in our 
case arise only for the $SU(3)$ group. Each of the stacks provides a $U(1)$ gauge group, and the
 weak hypercharge $Y$ is in general a linear combination of the U(1) charges $Q_a$ associated with the $a$th stack:
\begin{equation}
Y=c_1Q_1+c_2Q_2+\sum_ic_iQ_i+\sum_jc_jQ_j+\sum_kc_kQ_k
\end{equation}
For convenience we have suppressed any reference to any extra $U(1)$ factors 
which have been discussed after eq.(\ref{321p}). These are trivial to incorporate.

So we must also ensure that mixed $U(1)_Y-SU(3)_c$ and $U(1)_Y-SU(2)_L$ anomalies are absent, and their cancellation imposes further constraints 
on the wrapping numbers. To find them we must first fix the coefficients $c_a$.
The quark doublets $Q_L$ that arise from the $(12)$ intersection have $Q_1=1, \quad Q_2=-1$ and $Q_a=0, \quad a\neq 1,2$. Thus we require
\begin{equation}
c_1-c_2=\frac{1}{6}
\end{equation}
The $(1i)$   intersections, with $i \in I$, yield colour triplet, $SU(2)$ singlet fermions, which we 
require to be right-chiral  $u$ or $d$ quarks,  so 
\begin{equation}
c_1-c_i=\frac{2}{3}  \quad {\rm or} \quad -\frac{1}{3}
\end{equation} 
Similarly for the sets $J,K$. Thus we split the sets $I,J$ and $K$ into $I_{1,2}, J_{1,2}$ and $K_{1,2}$ with
\begin{eqnarray}
c_1-c_{i_1,j_1,k_1}&=&\frac{2}{3}\\
c_1-c_{i_2,j_2,k_2}&=&-\frac{1}{3}
\end{eqnarray} 
where $i_1 \in I_1$ etc. Then 
\begin{eqnarray}
Y&=&\left( c_1+\frac{1}{3} \right) \left( Q_1+Q_2+\sum_iQ_i+\sum_jQ_j+\sum_kQ_k \right) \nonumber \\
& -& \left( \frac{1}{3}Q_1+\frac{1}{2}Q_2+\sum_{i_1}Q_{i_1}+\sum_{j_1}Q_{j_1}+\sum_{k_1}Q_{k_1} \right)
\end{eqnarray}
It is easy to see that the combination $Q_1+Q_2+\sum_iQ_i+\sum_jQ_j+\sum_kQ_k$ is zero for all of the chiral fermions, so henceforth 
we take 
\begin{equation}
Y=-\left( \frac{1}{3}Q_1+\frac{1}{2}Q_2+\sum_{i_1}Q_{i_1}+\sum_{j_1}Q_{j_1}+\sum_{k_1}Q_{k_1} \right)
\end{equation}
Using (\ref{m}), the cancellation of the mixed $U(1)_Y-SU(3)_c$ anomaly then requires that the wrapping numbers satisfy
\begin{equation}
 \sum_{i_1}m_{i_1} + \sum_{j_1}m_{j_1} - 2\sum_{k_1}m_{k_1} =-3 \label{3anom}
\end{equation}
Chiral $SU(2)$ doublet fermions arise from the $I_{2a}=n_2m_a-3n_a$ intersections between the second stack and the $a$th stack, 
and cancellation of the mixed $U(1)_Y-SU(2)_L$ anomaly requires that the wrapping numbers also satisfy
\begin{eqnarray}
 n_2 \left( \sum_{i_1}m_{i_1} +1  - \sum_{k_1}m_{k_1} \right) 
\mbox{} + \sum_{j_1}n_{j_1}-2 \sum_{i_1}n_{i_1}+ \sum_{k_1}n_{k_1} =-1  
\end{eqnarray}
where we have used eqns.(\ref{n}),(\ref{m}) and (\ref{3anom}).

All intersecting branes give chiral fermions, but in addition tachyonic scalars arise whenever the Chan-Paton phases of 
the intersecting branes coincide. Thus we get $SU(2)_L$ doublet tachyons at $(2i_1)$ and $(2i_2)$
  intersections, charged singlet tachyons from $(i_1i_2)$,  $(j_1j_2)$ and $(k_1k_2)$ intersections, and
  $SU(3)_c$ colour triplet tachyons from $(1k_1)$ and $(1k_2)$ intersections. Doublet tachyons may be interpreted 
  as electroweak Higgs scalars, but the charged singlet and colour triplet tachyons are  less welcome. To avoid 
  the latter, the stacks in the sets $K_{1,2}$ must not intersect with those in the first stack, and this requires
  \begin{equation}
  m_{k_1}=0= m_{k_2} \quad \forall k_1 \in K_1, \quad \forall k_2 \in K_2
  \end{equation}
  Similarly, to avoid charged singlet tachyons we require
\begin{equation}
I_{i_1i_2}=0=I_{j_1j_2} \quad \forall i_1 \in I_1 , \quad \forall i_2 \in I_2, \quad \forall j_1 \in J_1 , \quad \forall j_2 \in J_2
\label{sixstack}
\end{equation}
As usual, the Yukawa couplings of the Higgs doublets generate fermion mass terms when the electroweak symmetry 
is spontaneously broken. The existence of Higgs doublets requires that
\begin{equation}
\sum_{i_1}|I_{2i_1}|+\sum_{i_2}|I_{2i_2}| \neq 0
\end{equation} 
 and we also demand that the model allows the 
(renormalizable) Yukawa couplings needed to give all quarks
 masses. This requires that
 \begin{eqnarray}
 I_1 \neq \emptyset \neq I_2 
\label{selection}
 \end{eqnarray}
Yukawa couplings arise from a disk-shaped world sheet with three open-string vertex operators attached to its boundary.
The boundaries of the disk are the relevant ${\rm D}4$-branes. For example, the insertion of a $Q_L$ vertex operator turns an $a=1$ stack
 ${\rm D}4$-brane boundary into a $b=2$ stack ${\rm D}4$-brane boundary \cite {AIQ}. 
The allowed Yukawas satisfy selection rules that derive from a $Z_2$ symmetry associated with each stack of ${\rm D}4$-branes.
 A state associated with a string 
between the $a$th and $b$th stack of ${\rm D}4$ branes is odd under 
the $a$th and $b$th $Z_2$ and even under any other $Z_2$.
It is straightforward to show that models with at most five stacks  cannot 
satisfy all of the above constraints.  
 The only six-stack models that do are parametrized by a single integer $n_2$ and are given in Table ~2.
 
 \begin{table}
\begin{center}
\begin{tabular}{|c|c|c|c|} \hline \hline
Stack $a$ & $N_a$ & $(n_a,m_a)$ & $\gamma_{\theta}$ \\
\hline \hline
1 & 3 & $(1,0)$ & $\alpha ^p {\bf I}_3$ \\
\hline
2 & 2 & $(n_2,3)$ & $\alpha ^q {\bf I}_2 $\\
\hline
$3=i_1$ & 1 & $(1-n_2,-3)$ & $\alpha ^q $\\
\hline
$4=i_2$ & 1 & $(1-n_2,-3)$ & $\alpha ^q $\\
\hline
$5=j_1$ & 1 & $(2,0)$ & $\alpha ^r$ \\
\hline
$6=k_1$ & 1 & $(-1,0)$ & $\alpha ^p$ \\
\hline \hline
\end{tabular}
\end{center}
\caption{Multiplicities, wrapping numbers and Chan-Paton phases for the six-stack models, 
($p\neq q \neq r \neq p$).}
\end{table}

The matter content of these models is easily determined using the results of \cite{AFIRU1}. In general, there 
are $n_G=3$ generations of  (massless) chiral fermions plus vector-like fermionic matter and 
(tachyonic) Higgs doublets of the form
\begin{equation}
\alpha (e^c+ \bar{e^c}) + \beta (L+\bar{L}) +\gamma_u(u^c+ \bar{u^c}) +\gamma_d(d^c+ \bar{d^c})+hH
\end{equation}
In our six-stack models this matter content turns out to be independent of $n_2$. We find
\begin{equation}
\alpha =3, \quad \beta = 9, \quad \gamma_u=0=\gamma_d, \quad h=6
\end{equation}
The gauge group is that of the standard model apart from some extra $U(1)$ factors 
provided $n_2\not = 0 \bmod 3$.
The weak hypercharge $Y$ is given by the superposition of $U(1)$ factors 
associated with the 6-stacks:
\begin{equation}
-Y=\frac{1}{3}Q_1+\frac{1}{2}Q_2+Q_3+(Q_5^{(1)}+Q_5^{(2)})+Q_6
\end{equation}
The mass of the tachyonic Higgs doublets \cite{AFIRU2} {\it does} depend on $n_2$ and is given by 
\begin{equation}
m_H^2=-\frac{m_s^2}{2\pi}\frac{\epsilon (n_2-\delta)}{|\delta||1-\delta|}
\label{mH2}
\end{equation}
where $m_s$ is the string scale; $\epsilon$ and $\delta$ are related to the parameters defining 
the torus wrapped by the D4-branes. If $R_1$ and $R_2$ are the radii of the two fundamental 
1-cycles, and $\theta$ is the angle between the two vectors defining the lattice, then
\begin{eqnarray}
\epsilon \equiv 2|\cos \theta/2| \label{eps}\\
\delta \equiv n_2-3R_2/R_1 \label{del}
\end{eqnarray}
The above formula for $m_H^2$ is valid so long as $\epsilon \ll 1$, but in any case $m_H^2 \ll m_s^2$ 
is required for consistency of the standard model without major contamination by string effects. $m_H^2$ 
also sets the scale for the various vector-like gonions that arise below the string scale. 
Specifically, we find the states whose masses and multiplicities are given in Table ~3. 
\begin{table}
\begin{center}
\begin{tabular}{|c|c|c|c|} \hline \hline
Intersection & Multiplicity  & Gonion & Mass$^2$ \\
\hline \hline
$(12)$ & 3 & Fermions $Q_L +\bar{Q}_L$ & $2n|1-\delta|m_H^2$ \\
\hline
$(13)$ & 3 & Fermions $u^c+ \bar{u^c}$ & $2n|\delta|m_H^2$\\
\hline
$(14)$ & 3 & Fermions $d^c+ \bar{d^c}$ & $2n|\delta|m_H^2$\\
\hline
$(23)$ & 6 & Fermions $L+\bar{L}$ & $2nm_H^2$\\
       & 6 & Scalars $H$ & $(2n+1)m_H^2$\\
       & 6 & Vectors $H$ & $(2n+1)m_H^2$\\
\hline
$(24)$ & 6 & Fermions $L+\bar{L}$ & $2nm_H^2$ \\
  & 6 & Scalars $H$ & $(2n+1)m_H^2$\\
       & 6 & Vectors $H$ & $(2n+1)m_H^2$\\
\hline
$(25)$ & 6 & Fermions $L+\bar{L}$ & $2n|1-\delta|m_H^2$ \\
\hline
$(26)$ & 3 & Fermions $L+\bar{L}$ & $2n|1-\delta|m_H^2$ \\
\hline
$(45)$ & 6 & Fermions $e^c+ \bar{e^c}$ & $2n|\delta|m_H^2$ \\
\hline
$(46)$ & 3 & Fermions $e^c+ \bar{e^c}$ & $2n|\delta|m_H^2$ \\
\hline \hline
\end{tabular}
\end{center}
\caption{Gonion multiplicities and masses. $n$ is an integer. } 
\end{table}

The matter content detailed above contributes to the running of the coupling strengths
 $\alpha_i(\mu) \quad (i=3,2,Y)$ 
  when    the renormalization scale $\mu$ is greater than the mass of the relevant matter, and we can hence 
  evaluate the coupling strengths at the (unknown) string scale $\mu=m_s$ at which their values are given in terms 
  of the Type II string coupling $\lambda_{II}$ and the wrapping numbers
   relevant to the stack producing the gauge group. We assume that the ${\mathcal N}=2$ gauge vector
supermultiplet is massless over the entire range starting from $\mu=m_Z$. 
There are also Kaluza-Klein and winding modes partners of gauge bosons. The 
former will have masses above the string scale $m_s$ provided 
\begin{equation}
\frac{\alpha_i(m_s)}{\lambda_{II}}\geq 1, \forall \;i
\label{KK}
\end{equation}
The latter will have masses above $m_s$ if in addition 
\begin{equation}
\frac{4\pi}{3} |\epsilon|(n_2-\delta)(m_s R_1)^2 \geq 1
\label{winding}
\end{equation} 
We shall assume eqns (\ref{KK}) and (\ref{winding}) in what follows.
In practice, we find values of $|\epsilon|\sim 10^{-1}$ so that 
Eq.(\ref{winding}) is satisfied when $R_1\geq m_s^{-1}$.
Then, at the string scale the gauge couplings are given by:
\begin{eqnarray}
\alpha_3^{-1}(m_s)&=&\alpha_3^{-1}(m_Z)+\frac{1}{\pi}\ln \frac{m_s}{m_Z} - \frac{1}{\pi}(N_3\ln X_3-\ln N_3!) \nonumber\\
 & & \mbox{} -\frac{1}{\pi}(N_2 \ln X_2 -\ln N_2!) \label{a3s}
\label{strong}
\end{eqnarray}
where
\begin{eqnarray}
X_3 \equiv \left| \frac{\delta \pi}{\epsilon (n_2 - \delta)} \right| \quad {\rm and} \quad N_3 \equiv [X_3] \\
X_2 \equiv \left| \frac{(1-\delta )\pi}{\epsilon (n_2 - \delta)} \right| \quad {\rm and} \quad N_2 \equiv [X_2]
\end{eqnarray}
The integers $N_{2,3}$ count the numbers of colour triplet gonions with masses below $m_s$. 
In eqn (\ref{strong}), and eqns (\ref{weak}), (\ref{hyper}) below, we have not included the effects of running the gonion and Higgs masses. 
We find that the gonion contributions to our solutions are small, so there is {\it post hoc} justification for our approximation. 
\begin{eqnarray}
\alpha_2^{-1}(m_s)&=&\alpha_2^{-1}(m_Z)-\frac{6}{2\pi}\ln \frac{m_h}{m_Z}
-\frac{7}{2\pi}\ln \frac{m_s}{m_h} - \frac{3}{\pi}(N_3\ln X_3-\ln N_3!) \nonumber\\
 & & \mbox{} -\frac{2}{\pi}(N_1 \ln X_1 -\ln N_1!) -\frac{3}{4\pi}\left[ (N_0+1)\ln 2X_1 -\ln \frac {(2N_0+1)!}{2^{N_0}N_0!} \right] \label{a2s}
\label{weak}
\end{eqnarray}
where
\begin{eqnarray}
X_1 \equiv \left| \frac{\delta (1-\delta) \pi}{\epsilon (n_2 - \delta)} \right|,  \quad N_1 \equiv [X_1]  
\quad {\rm and} \quad N_0 \equiv [X_1-1/2]
\end{eqnarray}
$N_1$ counts the number of lepton doublet gonions below the string scale, and $2N_0+1$ is the number of scalar and vector 
gonions with masses below this scale.
\begin{eqnarray}
\alpha_Y^{-1}(m_s)&=&\alpha_Y^{-1}(m_Z)-\frac{50}{6\pi} \ln\frac{m_h}{m_Z}
-\frac{53}{6\pi}\ln \frac{m_s}{m_h} - \frac{5}{3\pi}(N_3\ln X_3-\ln N_3!) \nonumber\\
\mbox{} -\frac{14}{3\pi}(N_2 \ln X_2 -\ln N_2!) & \mbox{} - & \frac{2}{\pi}(N_1 \ln X_1 -\ln N_1!) \nonumber \\ 
& \mbox{}- &\frac{3}{4\pi}\left[ (N_0+1)\ln 2 X_1 -\ln \frac {(2N_0+1)!}{2^{N_0}N_0!} \right] \label{aYs}
\label{hyper}
\end{eqnarray}
The ratios of the gauge coupling strengths, 
which depend on the length of the cycles 
$(n_a,m_a)$ and on the superposition of the 
$U(1)$ factors in the weak hypercharge  \cite{AFIRU1},  are  
independent of the unknown $\lambda_{II}$: 
\begin{eqnarray}
\frac{\alpha_2^{-1}(m_s)}{\alpha_3^{-1}(m_s)}&=&\left[ \delta^2+n_2(n_2-\delta)\epsilon^2 \right]^{1/2} \label{a23}\\
\frac{\alpha_Y^{-1}(m_s)}{\alpha_3^{-1}(m_s)}&=&\frac{10}{3} +\frac{\alpha_2^{-1}(m_s)}{2\alpha_3^{-1}(m_s)}+
\left[ (1-\delta)^2+(n_2-1)(n_2-\delta)\epsilon^2 \right]^{1/2} \label{aY3}
\end{eqnarray}
where $\epsilon$ and $\delta$ are defined in equations (\ref{eps}) and (\ref{del}). Thus, by substituting the solutions (\ref{a3s}),(\ref{a2s}) 
and (\ref{aYs}) of the renormalization group equations into (\ref{a23}) and (\ref{aY3}), we obtain two constraint equations on  
$\epsilon$ and $\delta$. 

We also considered the case when the ${\mathcal N}=2$ 
supersymmetric partners of the gauge fields acquire masses on the scale of 
$m_s$ or greater so they do not contribute to the renormalization group 
equations. In that case,  the coefficient of the 
$\rm{ln}\frac{m_s}{m_Z}$ in eqn.(\ref{strong}) is replaced by $\frac{7}{2\pi}$, 
the coefficients of ${\rm ln}\frac{m_h}{m_Z}$ and ${\rm ln}\frac{m_s}{m_h}$ 
in eqn.(\ref{weak}) become $-\frac{4}{3\pi}$ and $-\frac{11}{6\pi}$ 
respectively, and in eqn.(\ref{hyper}) become $-\frac{25}{3\pi}$ and 
$-\frac{53}{6\pi}$ respectively.

We solve the constraints (\ref{a23}) and 
(\ref{aY3}) numerically for a range of values of 
the parameters $n_2$ and $m_s$, 
and using the latest values \cite{cernyellow} of the standard model parameters which yield  
\begin{eqnarray}
\alpha_3^{-1}(m_Z)=8.403 \\
\alpha_{2}^{-1}(m_Z)=29.776 \\
\alpha_{Y}^{-1}(m_Z)=99.124
\end{eqnarray}
for  the coupling strengths at the weak scale. For  given values of the parameters $n_2$ and $m_s$, 
the calculated values of $\epsilon$ and $\delta$
determine the mass of the tachyonic Higgs doublets using equation (\ref{mH2}), and this determines the mass 
$m_h$ of the physical Higgs particle from
\begin{equation}
m_h^2=|2m_H^2|
\end{equation}
The ratio $a$ of the compactification radii is then determined from  eqn (\ref{del}).  Our results are summarised in Table ~4 
for the case of an ${\mathcal N}=2$ supersymmetric gauge sector, and in Table ~5 when the ${\mathcal N}=2$ gauge superpartners are assumed to 
have masses of order $m_s$. 

\begin{table}[h]
\begin{center}
\begin{tabular}{|c|c|c|c|c|} \hline \hline
 $n_2$ & $m_s \quad ({\rm TeV})$ & $\epsilon$ & $m_h \quad ({\rm GeV})$ & $a$ \\
\hline \hline
5 & 1.0 & 0.185 & 137 & 0.294 \\
\hline 
5 & 1.2 & 0.182 & 165 & 0.297 \\
 \hline
5 & 1.9 & 0.176 & 265 & 0.304 \\
\hline 
5 & 3.0 & 0.170 & 425 & 0.312 \\
\hline \hline
10 & 1.0 & 0.123 & 137 & 0.198 \\
\hline 
10 & 1.2 & 0.121 & 165 & 0.199 \\
\hline 
10 & 1.9 & 0.116 & 265 & 0.202 \\
\hline 
10 & 3.0 & 0.111 & 425 & 0.206\\ 
\hline \hline
20 & 1.0 & 0.074 &  137 & 0.119\\
\hline
20 & 1.2 & 0.073 & 165 & 0.120 \\
\hline 
20 & 1.3 & 0.072 & 179 & 0.120 \\
\hline 
20 & 1.9 & 0.069 & 265 & 0.121 \\
\hline 
20 & 3.0 & 0.066 & 425 & 0.122 \\
\hline \hline
\end{tabular}
\end{center}
\caption{Predicted values of the physical Higgs mass $m_h \equiv |2m_H^2|^{1/2}$ and the ratio  $a \equiv R_1/R_2 $
 of the compactification radii  in the case of an ${\cal N}=2$ supersymmetric gauge sector.}
\end{table}

 \begin{table}[h]
\begin{center}
\begin{tabular}{|c|c|c|c|c|} \hline \hline
 $n_2$ & $m_s \quad ({\rm TeV})$ & $\epsilon$ & $m_h \quad ({\rm GeV})$ & $a$ \\
\hline \hline
5 & 1.0 & 0.130 & 145 & 0.345 \\
\hline 
5 & 1.2 & 0.125 & 175 & 0.353 \\
 \hline
5 & 1.9 & 0.112 & 286 & 0.370 \\
\hline 
5 & 3.0 & 0.099 & 466 & 0.389 \\
\hline \hline
10 & 1.0 & 0.082 & 145 & 0.219 \\
\hline 
10 & 1.2 & 0.078 & 175 & 0.222 \\
\hline 
10 & 1.9 & 0.069 & 286 & 0.229 \\
\hline 
10 & 3.0 & 0.06 & 466 & 0.236\\ 
\hline \hline
20 & 1.2 & 0.045 & 175 & 0.128 \\
\hline 
20 & 1.3 & 0.044 & 191 & 0.128 \\
\hline 
20 & 1.9 & 0.039 & 286 & 0.130 \\
\hline 
20 & 3.0 & 0.034 & 466 & 0.132 \\
\hline \hline
\end{tabular}
\end{center}
\caption{Predicted values of the physical Higgs mass $m_h \equiv |2m_H^2|^{1/2}$ and the ratio  $a \equiv R_1/R_2 $
 of the compactification radii in the case when the ${\cal N}=2$ gauge superpartners have masses of order $m_s$.}
\end{table}

We find that it is easy to obtain values of $\epsilon \ll 1$ and $m_H^2 \ll m_s^2$ consistently with a string scale $m_s$ 
not more than a few TeV. Physical Higgs masses not much greater than the current LEP bound ($m_h>114$ GeV) \cite{LEPmH} are found for some 
values of the parameters, but these are probably excluded because they are necessarily accompanied by charged, vector-like
 gonions with similar masses. The general behaviour is that with $m_s$ fixed, $m_h$ falls slowly as $n_2$ increases, whereas with $n_2$ fixed,
 $m_h$ increases as $m_s$ increases.
As in  \cite{AIQU,BKL}, with ${\rm D}$-branes located at $R/Z_3$ orbifold 
fixed points, the mixed $U(1)$ anomalies in the present intersecting brane model are cancelled by a generalized 
Green-Schwarz mechanism mediated by twisted Ramond-Ramond fields. However, it is unclear whether the anomalous $U(1)$s survive as 
global symmetries because of the lack of supersymmetry in the present set-up.
Certainly, the Yukawa terms needed to generate the required masses conserve all of the $U(1)$ symmetries, so it seems likely that they 
are global symmetries. If so, keV-scale axions will remain a problem in these models too.
Unlike the models of \cite{AIQU},\cite{BKL}, the present model possesses 
lepton mass terms as well as quark mass terms, at renormalizable level 
provided we assign the lepton doublets to (25) interesections rather than the (24) 
intersection. Also unlike these models our model is free from lepton number violating terms at renormalizable level. On the other hand it shares with those models 
the virtue of proton decay being perturbatively forbidden as a result of the 
selection rules discussed after (\ref{selection}) \cite{AFIRU2}, although baryon number $B$ non-conservation with $\Delta B$ even {\it is} allowed. 
Masses for the vector-like matter are more problematic because of the lack of 
gauge-singlet scalars amongst the tachyonic states to develop vacuum expectation values. 
Three of the $L+\bar{L}$ states and three of the $e_L^c+{\bar {e}}_L^c$ states acquire masses at tree-level via the usual $Le_L^c H$ and $\bar{L}{\bar e}_L^c
H^{\dagger}$ couplings. The remaining six $L+\bar{L}$ pairs are able to acquire masses via non-renormalizable terms of the form $L\bar{L}HH^{\dagger}$; the latter 
are suppressed by a factor $<H>/m_s$.

In conclusion, we have found a unique class of six-stack standard-like models that give masses to all of the matter, and are 
free of charged singlet  (and colour triplet) tachyons. We find it relatively easy to ensure that the Higgs mass is small 
compared with the $1-3$ TeV  string scale, while consistently reproducing the observed values of the gauge coupling constants 
at the electroweak scale, and achieving the values required by the string theory at the string scale. However, our models {\it do} have extra light, 
vector-like lepton states (but not quark states), and Higgs doublets, as well as the towers of gonions characteristic of all theories of this type. There are also three 
extra non-anomalous $U(1)$ gauge groups under which the matter is charged.

\newpage

\section*{Acknowledgements}
This research is supported in part by PPARC. We are grateful to Angel Uranga for pointing out the error in the previous version of this 
paper concerning the generalized Green-Schwarz mechanism.

\end{document}